\documentclass[11pt]{article}
\usepackage{xspace}
\usepackage{graphicx}
\usepackage{amsmath}
\usepackage{amssymb}
\usepackage{color}
\definecolor{Red}{rgb}{1,0,0}
\definecolor{Green}{rgb}{0,1,0}
\definecolor{Blue}{rgb}{0,0,1}
\definecolor{Black}{rgb}{0,0,0}

\def\beq{\begin{equation}}
\def\eeq#1{\label{#1}\end{equation}}
\def\eeqn{\end{equation}}

\def\beqa{\begin{eqnarray}}
\def\eeqa#1{\label{#1}\end{eqnarray}}
\def\eeqan{\end{eqnarray}}

\let\bar=\overbar

\def\Dslash{\not{\hbox{\kern-4pt $D$}}}
\def\dslash{\not{\hbox{\kern-2pt $\del$}}}

\def\msb{{\bar{\ssstyle M \kern -1pt S}}}

\def\Title#1{\begin{center} {\Large {\bf #1} } \end{center}}

\begin{document}

\Title{nuSTORM: Neutrinos from Stored Muons}

\bigskip\bigskip


\begin{raggedright}  

{\it F.J.P. Soler\footnote{On behalf of the nuSTORM Collaboration}\index{Soler, F.J.P.},\\
School of Physics and Astronomy,\\ 
University of Glasgow,\\
Glasgow, G12 8QQ, UK}\\

\end{raggedright}
\vspace{1.cm}

{\small
\begin{flushleft}
\emph{To appear in the proceedings of the Prospects in Neutrino Physics Conference, 15 -- 17 December, 2014, held at Queen Mary University of London, UK.}
\end{flushleft}
}


\section{Introduction}
Decay rings for muons were proposed by Koshkarev in 1974 \cite{Koshkarev:1974my} and Neuffer in 1980 \cite{Neuffer:1980ru}.
This led to developments of the modern concept of a Neutrino Factory, which may be seen as a first step towards a Muon Collider. 
The physics capabilities of a Neutrino Factory were described by Geer \cite{Geer:1997iz}.
The neutrino beam is created from the decay of muons in flight in a
storage ring, and may be used to discover CP violation in the neutrino sector and to resolve the neutrino mass ordering.  
Since both $\mu^+$ and $\mu^-$ can be
created with the same systematic uncertainties on the flux, any
oscillation channel can be studied with both neutrinos and
antineutrinos, improving sensitivity to CP
violation. 

A simple first step towards a Neutrino Factory could be the nuSTORM (Neutrinos from STORed Muons) decay ring. 
The main motivation for nuSTORM is to design and construct a new type of neutrino beam with well-understood characteristics from 
decays of muons that does not require any new  technology, such as ionisation cooling. 
The facility consists of a 3.8\,GeV/c muon storage ring that can be used to study eV-scale neutrino oscillation physics,
$\nu_e$ and $\nu_\mu$ interaction physics and to develop technology for future accelerator projects. The facility can 
search for sterile neutrinos in both appearance and disappearance modes,  provide precise studies of electron and
muon neutrino scattering on nuclei, in an energy appropriate for future long- and short-baseline neutrino oscillation programmes, and
provide the technology test-bed required to carry-out the R\&D critical for the implementation of the next step in a
muon-accelerator based particle-physics programme.

\section{The nuSTORM concept}
\begin{figure}[htpb]
  \centering{
    \includegraphics[width=0.8\textwidth]{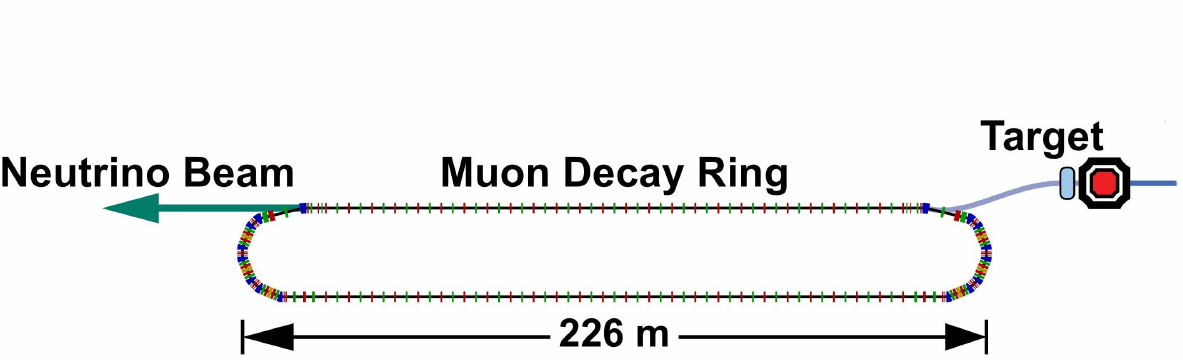}}
  \caption{A schematic of the storage ring configuration. Pions are
      injected into a straight section and must decay into muons before the
      first bend or be ejected from the ring. Muons that decay in the
      injection straight during subsequent turns produce the neutrino beam.}
  \label{fig:nuSTORM}
\end{figure}

The  nuSTORM facility consists of a ring designed to store muons with momenta of 3.8\,GeV/c. Protons of 120\,GeV/c are used to produce pions from a 
conventional solid target.  The pions are collected with a magnetic horn and quadrupole magnets
and those with 5\,GeV/c momentum are injected into a storage ring.  The pions that decay in the first straight of the ring yield muons, and those with
momenta of 3.8\,GeV $\pm$ 10\% are captured in the storage ring. The circulating muons then 
subsequently decay into electrons and neutrinos.  The momentum was selected to maximise the physics reach for both 
neutrino oscillations and to measure neutrino cross sections in an energy relevant for long-baseline neutrino oscillation experiments.  
See Figure \ref{fig:nuSTORM} for a schematic of the facility. More details of the nuSTORM facility may be found in References 
 \cite{Kyberd:2012iz,Adey:2013pio,Adey:2013afh}.
 
\subsection{Physics motivation of nuSTORM}
A number of results have been reported that can be interpreted as
hints for oscillations involving sterile neutrinos with masses at the eV scale
(for a recent review see \cite{Abazajian:2012ys}).
nuSTORM is capable of making the measurements required to confirm
or refute the evidence for sterile neutrinos using a
technique that is both qualitatively and quantitatively new.
The nuSTORM facility delivers beams of $\nu_e$
($\bar{\nu}_e$) and $\bar{\nu}_\mu$ ($\nu_\mu$).
A detector located at a distance $\sim 2\,000$\,m from the end of one
of the straight sections will be able to make sensitive searches for
the existence of sterile neutrinos.
If no appearance ($\nu_e \rightarrow \nu_\mu$)
signal is observed, the allowed region can be ruled out at the
$\sim 10 \sigma$ level \cite{Adey:2014rfv}.
Instrumenting the nuSTORM neutrino beam with a near detector at a
distance of $\sim 20$\,m makes it possible to search for sterile
neutrinos in the disappearance $\nu_e \rightarrow \nu_X$ and
$\bar{\nu}_\mu \rightarrow \bar{\nu}_X$ channels.
In the disappearance search, the absence of a signal would permit the 
presently allowed region to be excluded at the 99\% confidence level. 
For a general discussion of the optimisation of disappearance searches at short baselines
see \cite{Winter:2012sk}.

With the discovery  that $\theta_{13} $ is non-zero \cite{An:2012eh,Ahn:2012nd,Abe:2011fz,Abe:2011sj,Adamson:2011qu} 
and that it has a larger than expected value ($\sin^2 2 \theta_{13} \sim 0.1$),
the search for CP violation in the lepton sector requires measurements of oscillation
probabilities with uncertainties at the percent level.
For future long-baseline experiments to reach their
ultimate precision requires that the $\nu_e N$ and
the $\nu_\mu N$ cross sections are known precisely for
neutrino energies ($E_\nu$) in the range $0.5 < E_\nu < 3$\,GeV.  
nuSTORM is unique as it makes it possible to measure the
$\bar{\nu}_\mu$ ($\nu_\mu$) and $\nu_e$ ($\bar{\nu}_\mu$) nucleus cross 
sections with a precision $\simeq 1$\% over the required
neutrino-energy range. The flavour composition of the beam and the
neutrino-energy spectrum are both precisely known.
The storage-ring instrumentation combined with measurements at a
near detector will allow the neutrino flux to be determined with a flux accuracy of $\simeq 10^{-3}$, 
with the potential to transform the field of neutrino interaction physics.

Finally, the unique capabilities of nuSTORM  offer the opportunity to provide muon beams 
for future investigations into six dimensional muon ionisation cooling, while running the neutrino programme simultaneously.
Muon cooling is the key enabling technology needed for future ultra-high intensity muon accelerator facilities.
Its demonstration would be one of the major steps towards the realisation of a multi-TeV Muon Collider.

\subsection{Long baseline physics}
Neutrino Factories have demonstrated the best potential sensitivity for CP violation of all future long-baseline neutrino oscillation facilities.
To measure the CP asymmetry ($A_{CP}$) at $3\sigma$ for 75\% coverage of the values of $\delta_{CP}$, means that $A_{CP}$ may be as low as 5\%, which
requires a precision of 1.5\% and a systematic error of order  1\%. However, we measure the rate: 
\begin{equation}
R_{\alpha\beta}\left(E_{vis}\right) =  N\int dE \Phi_\alpha (E)\sigma_\beta \left(E,E_{vis}\right)\epsilon_\beta\left(E\right)P\left(\nu_\alpha \rightarrow \nu_\beta, E\right),
\end{equation}
where $E$ is the neutrino energy, $E_{vis}$ is the visible energy, $\Phi_\alpha (E)$ is the flux of $\nu_\alpha$, $\sigma_\beta \left(E,E_{vis}\right)$ is the cross section
and $\epsilon_\beta$ is the efficiency of detection of a $\nu_\beta$ of energy $E$ but measured at a visible energy $E_{vis}$, and 
$P\left(\nu_\alpha \rightarrow \nu_\beta, E\right)$ is the probability of oscillation from $\nu_\alpha$ to $\nu_\beta$ \cite{Huber:2014nga}.
In a disappearance experiment, we can satisfy:
\begin{equation}
\frac{R_{\alpha\beta}\left( far\right) L^2}{R_{\alpha\beta}\left( near\right)} \sim  \frac{N_{far} \Phi_\alpha \sigma_\beta \epsilon_\beta P\left(\nu_\alpha \rightarrow \nu_\beta \right)}{N_{near} \Phi_\alpha \sigma_\alpha \epsilon_\alpha},
\end{equation}
where $\alpha = \beta$, so many of the systematic errors cancel. However, in an appearance experiment,   $\alpha \neq \beta$, so the $\nu_\alpha$ beam cannot be used to measure $\sigma_\beta \epsilon_\beta$. The difference in the $\sigma_{\nu_e}$ and $\sigma_{\nu_\mu}$ cross sections can be large, depending on the energy, due to nuclear effects and this may affect the neutrino oscillation measurements. For example, in the Tokai to Hyper-Kamiokande (T2HK) proposed experiment, 
with a precision in the ratio of cross sections $\sigma_{\nu_e}/\sigma_{\nu_\mu}$ between 1\% and 2\% one can achieve a 3$\sigma$ measurement of CP violation for 75\%
of the $\delta_{CP}$ values with an exposure between 600 and 800 kton$\cdot$MW$\cdot$years. However, a degradation of the systematic uncertainty to the 5\% level corresponds to an increase in the exposure of roughly 200-300\% required to achieve 3$\sigma$ accuracy for 75\% of the $\delta_{CP}$ values  \cite{Huber:2007em}.
Muon storage rings, such as nuSTORM or a Neutrino Factory, are the only known facilities that can achieve a measurement of cross sections with less than 1\% precision.

\subsection{Short baseline physics}
The LSND \cite{Athanassopoulos:1996jb,Aguilar:2001ty} and MiniBooNE experiments \cite{AguilarArevalo:2010wv,Aguilar-Arevalo:2013pmq} show hints of  
$\bar{\nu}_\mu\rightarrow \bar{\nu}_e$ and $\nu_\mu\rightarrow \nu_e$ appearance,  that can be explained by an oscillation mediated by sterile neutrino states, in short-baseline accelerator experiments. Furthermore, there is additional evidence of a 6\% deficit of $\bar{\nu}_e$ from reactor
experiments (the reactor anomaly) based on more accurate recent re-evaluations of the reactor antineutrino flux 
\cite{Mueller:2011nm,Mention:2011rk,Huber:2011wv,Abe:2011fz} that can also be interpreted as antineutrino disappearance via sterile neutrino mediated oscillations.
Finally, active-to-sterile neutrino oscillations can also explain the gallium anomaly, in which intense artificial radioactive sources used to calibrate gallium radiochemical detection experiments observed fewer neutrinos from the source than expected \cite{Acero:2007su,Giunti:2012tn}. 

Global fits attempt to explain these data, but there exists tension between the appearance and disappearance measurements \cite{Kopp:2013vaa}.  The sterile neutrino hypothesis is satisfied when:
\begin{equation}
P\left(\nu_\mu \rightarrow \nu_e \right) \leq 4 \left(1-P\left(\nu_\mu \rightarrow \nu_\mu \right)\right)\left(1-P\left(\nu_e \rightarrow \nu_e \right)\right).
\end{equation}
The nuSTORM facility could probe all possible sterile neutrino appearance  and disappearance channels to test the sterile neutrino paradigm in detail.

\section{nuSTORM parameters}
The nuSTORM facility is designed to produce 3.8\,GeV/c muons that are injected and stored in a storage ring (Figure \ref{fig:nuSTORM}). A 100\,kW proton beam of 120\,GeV energy impinges on a carbon or an inconel target. Pions produced in the target are captured in a NuMI-style horn, they are then transported down a transfer line and 5\,GeV/c ($\pm 20\%$) pions are stochastically injected into a storage ring. The target, collection system and stochastic injection systems have been designed to deliver 0.11 pions per proton on target (POT) \cite{Kyberd:2012iz} to the storage ring.  

\begin{figure}[htbp]
    \centering{
    \includegraphics[width=0.48\columnwidth]{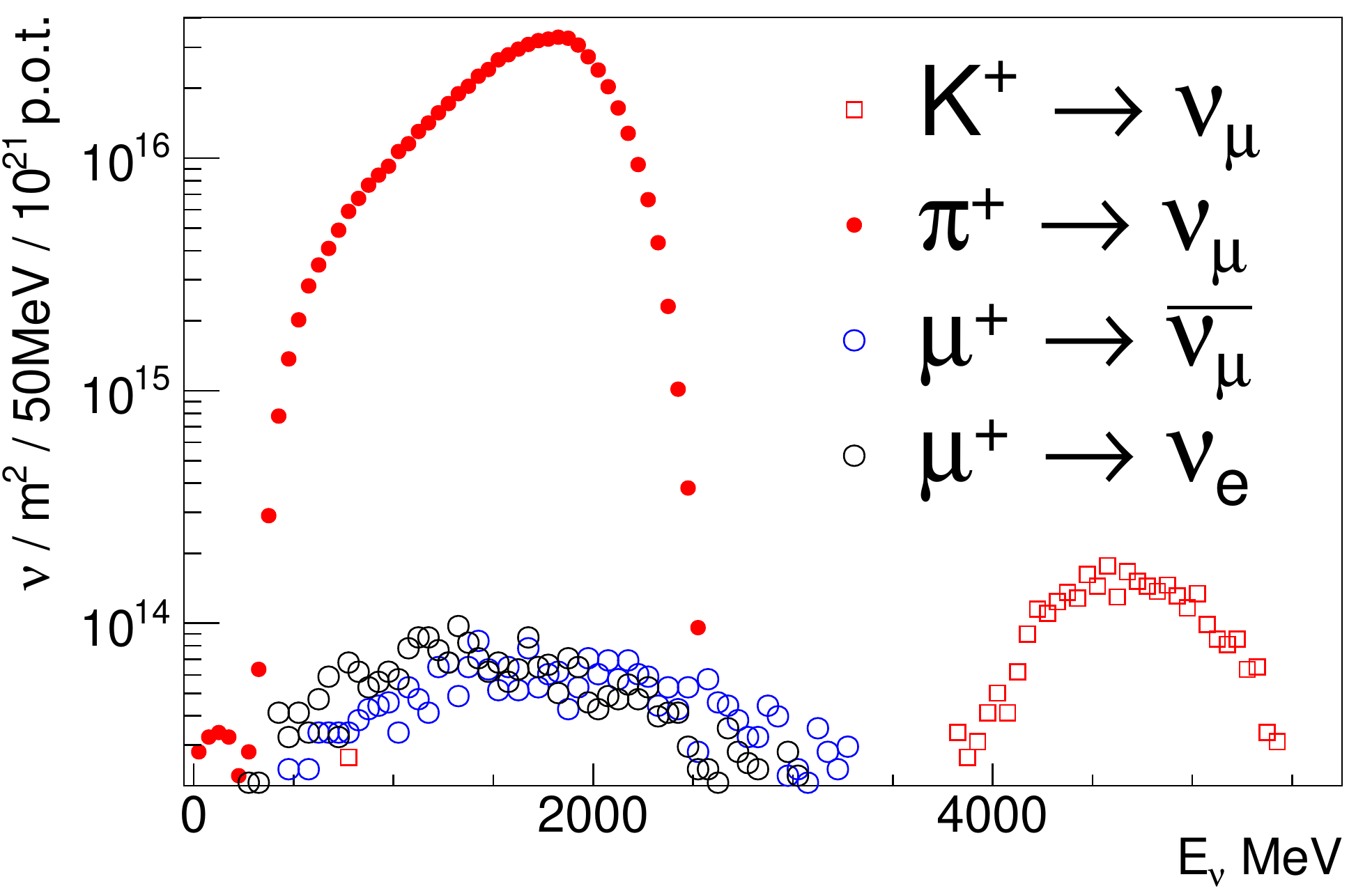}
      \includegraphics[width=0.51\columnwidth]{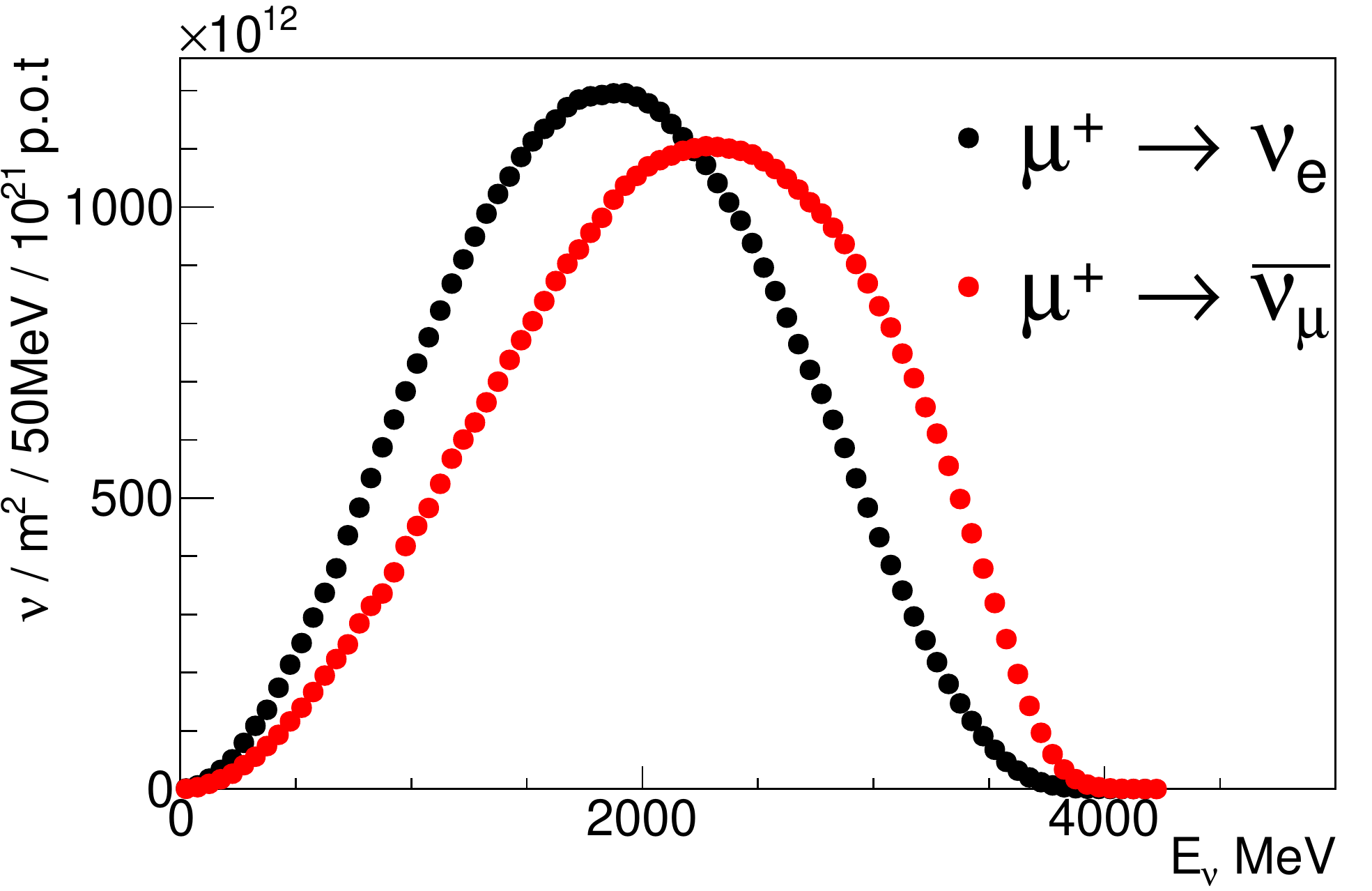}}
    \caption{nuSTORM flux and energy spectrum from the pion flash just after injection (left) and from muon decay over 100 turns (right).}
    \label{fig:spectrum}
\end{figure}

The storage ring consists of a large aperture FODO  lattice designed to transport muons of 3.8\,GeV/c ($\pm$ 10\%) momenta around the ring. It is 
calculated that 52\% of pions decay to muons before the first turn and $8\times 10^{-3}$ muons per POT are stored in the storage ring. For $10^{20}$ POT, 
we expect a flash of neutrinos from $8.6\times 10^{18}$ pion decays and 
we expect $2.6\times 10^{17}$  positive muons that decay in the ring (the muon lifetime is 27 orbits of the decay ring). 
The nuSTORM flux and energy spectrum from the pion flash and from recirculating muons are shown in Figure \ref{fig:spectrum}. 

The flux of $\nu_\mu$ from pion decays is $6.3\times 10^{16}$~$\nu$/m$^2$, the flux of 
$\nu_e$  from muon decays is $3.0\times 10^{14}$~$\nu$/m$^2$ and $\nu_\mu$ from kaon decay is $3.8\times 10^{14}$~$\nu$/m$^2$, all at a distance of 50~m. 
This hybrid beam from pion and muon decay can produce a rich physics programme of neutrino cross-section measurements and can be used to perform a 
sterile neutrino search at a short-baseline oscillation experiment. Furthermore, the flux uncertainties for nuSTORM are less than 1\%, due to the precise knowledge of
the muon decay spectrum and from the instrumentation that can be installed in the storage ring to measure the number of muons in the ring. 
The event rates per $10^{21}$ POT in 100 tons of a Liquid Argon detector at 50~m are shown in Table \ref{tab:rates}. 
\begin{table}[htbp]
\begin{center}
\caption{Event rates per $10^{21}$ POT in 100 tons of Liquid Argon at 50~m from the nuSTORM storage ring.  }
\label{tab:rates}
	\begin{tabular}{|cc|cc|}
	\hline
	 $\mu^+$ Channel & $N_{evts}$ & $\mu^-$ Channel & $N_{evts}$ \\
	\hline
	$\bar{\nu}_\mu$ NC & 1,174,710 & $\bar{\nu}_e$ NC & 1,002,240\\
	$\nu_e$ NC & 1,817,810 & $\nu_\mu$ NC & 2,074,930\\
	$\bar{\nu}_\mu$ CC & 3,030,510 & $\bar{\nu}_e$ CC & 2,519,840\\
	$\nu_e$ CC & 5,188,050 & $\bar{\nu}_\mu$ CC & 6,060,580\\
	\hline
	$\pi^+$ Channel & $N_{evts}$ & $\pi^-$ Channel & $N_{evts}$ \\
	\hline
	$\nu_\mu$ NC & 14,384,192 & $\bar{\nu}_\mu$ NC & 6,986,343\\
	$\nu_\mu$ CC & 41,053,300 & $\bar{\nu}_\mu$ CC & 19,939,704\\
	\hline
	\end{tabular}
\end{center}
\end{table}

\section{Neutrino interaction physics}
There is a very rich physics programme in neutrino interaction physics that may be performed at the nuSTORM facility, due to its very large event rate and its 
accurate flux and energy determination. It will be able to perform crucial $\nu_\mu$, $\nu_e$, $\bar{\nu}_\mu$ and $\bar{\nu}_e$ cross-section measurements, 
required for the long-baseline neutrino oscillation programmes, with statistical and systematic uncertainties of less than 1\%, using the storage-ring instrumentation.

\begin{figure}[h]
 \begin{center}
  \includegraphics[width=0.7\textwidth]{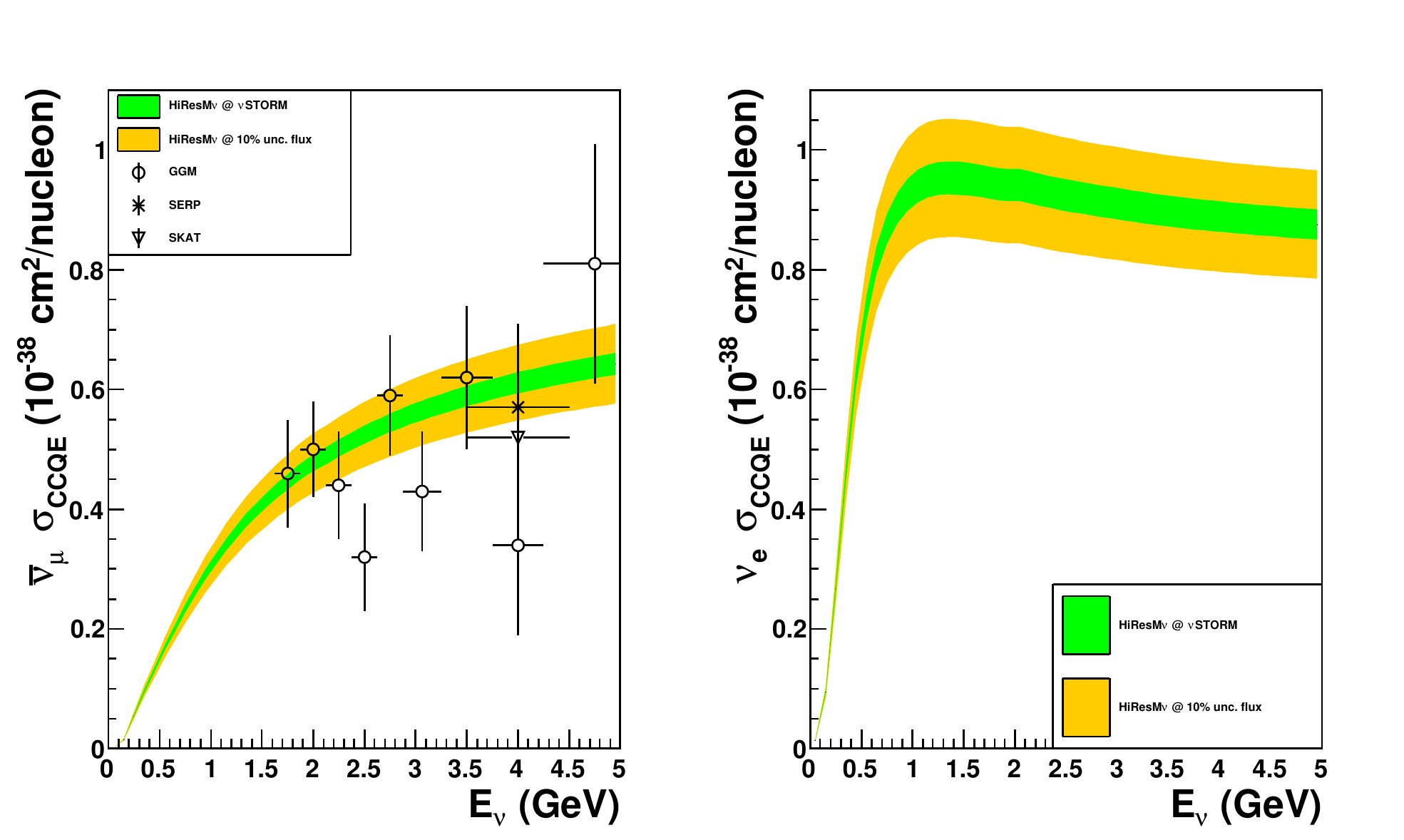} 
  \includegraphics[width=0.7\textwidth]{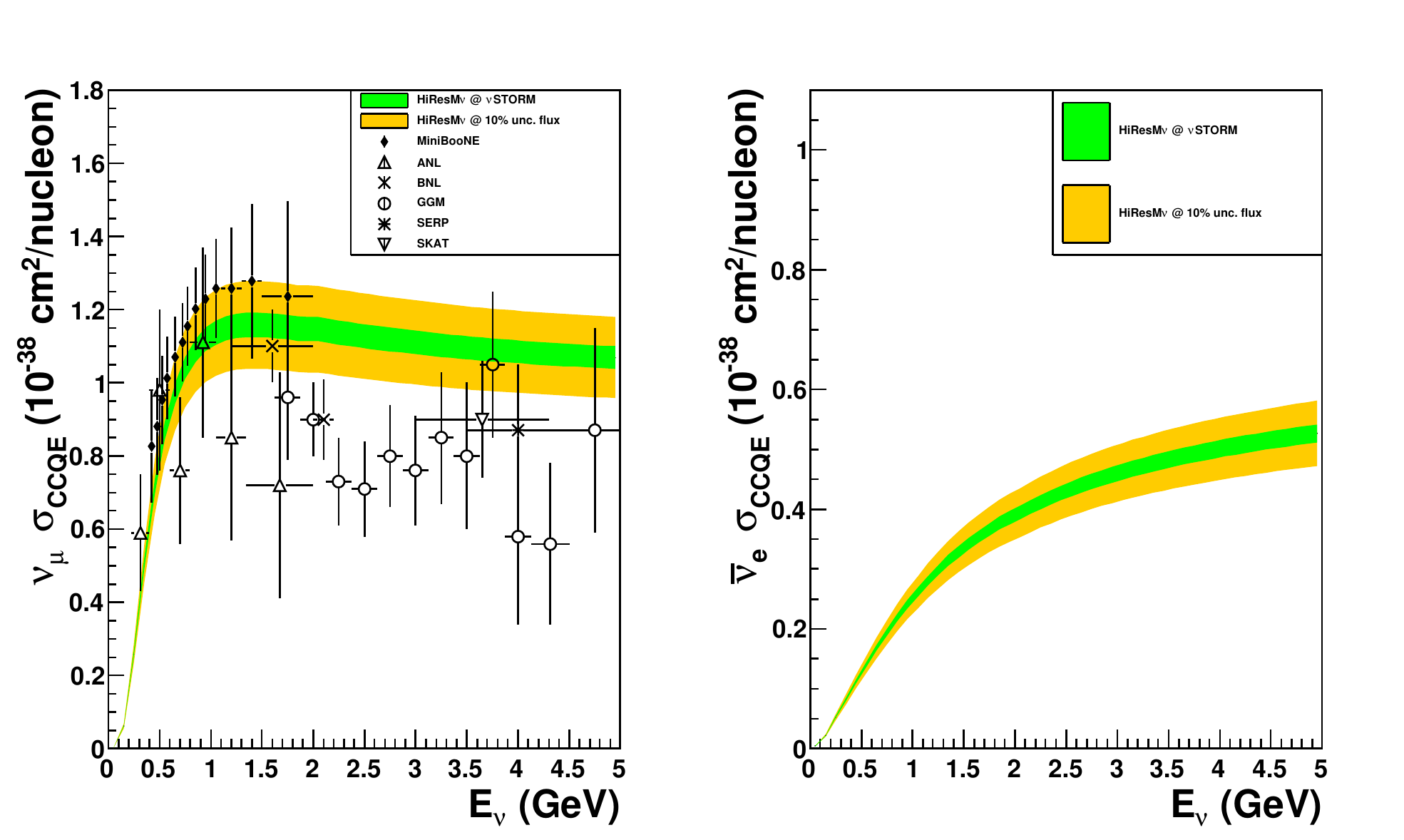}
 \end{center}
\caption{
 The CCQE cross sections ($\sigma_{\rm CCQE}$) plotted as a function
  of incident neutrino energy ($E_\nu$) for $\bar{\nu}_\mu$ (top, left), $\nu_e$ (top, right), $\nu_\mu$  (bottom, left) and $\bar{\nu}_e$ (bottom, right)
  interactions with the HiResM$\nu$ detector at nuSTORM (the green band includes 1\% flux uncertainty and the yellow band shows 10\% flux uncertainty). 
}
\label{genie_ccqe}
\end{figure}

A number of near detectors are being considered for the facility, such as a detector similar to the HiResM$\nu$ detector proposed for DUNE at LBNF \cite{Mishra:2008nx}.
A high pressure gaseous or a liquid argon time projection chamber (TPC) would also be suitable choices. The performance of the HiResM$\nu$ 
detector exposed to the nuSTORM flux was studied 
in \cite{Adey:2013pio}. The charged-current quasi-elastic (CCQE) cross sections are plotted as a function of neutrino energy in Figure \ref{genie_ccqe}. 
The figure shows the precision with which the cross sections would be measured if the systematic uncertainties estimated for the HiResM$\nu$ detector are combined with the 1\% flux uncertainty that nuSTORM will provide, compared to a flux uncertainty of 10\%. Figure  \ref{genie_ccqe} also shows the present measurements of the CCQE cross sections  (only available for muon-neutrino and muon-anti-neutrino beams). The nuSTORM facility has the potential to improve the systematic uncertainty on $\nu_\mu$ and $\bar{\nu}_\mu$  CCQE cross section measurements by a factor of  5 -- 6.  It is only very recently that preliminary results on $\nu_e$ CCQE cross section measurements are available \cite{Wolcott:2015lya} so nuSTORM will be able to perform comprehensive measurements with both $\nu_e$  and $\bar{\nu}_e$ beams.

Further to the potential to perform world-leading CCQE measurements, nuSTORM will probe other neutrino scattering topics, such as $\pi^0$ production in neutrino interactions, charged pion and kaon production, inclusive charged current and neutral currents, including the ratio of these, which can be used to determine
 $\sin^2\theta_W$.  It will be able to probe nuclear effects in neutrino interactions and other semi-exclusive and exclusive processes such as measurements of $K_s$, 
 $\Lambda$ and $\bar{\Lambda}$ production. The nuSTORM facility can be used to search for new physics effects and exclusive processes, such as  tests of  $\nu_\mu$ and $\nu_e$ universality, searches for heavy neutrinos and eV-scale pseudo-scalar penetrating particles. In summary, there exists a very rich physics programme
to be carried out at a near detector at the nuSTORM facility that has the potential to revolutionise neutrino interaction physics.
  
 \section{Sterile neutrino search}
With the event rates expected at nuSTORM, one can carry out a sterile neutrino search in a short-baseline experiment, with a far detector, consisting of a 1.3 kton magnetised iron 
detector (Super BIND) at 2~km, and a toroidal magnetic field between 1.5-2.6~T, fed by a superconducting transmission line delivering 240~kA-turns. A neutrino oscillation
experiment can be carried out at the nuSTORM facility in which one can carry out  simultaneously a $\nu_\mu$ appearance search and a  $\bar{\nu}_\mu$ disappearance search. The probability of observing a $\nu_{e}\to\nu_{\mu}$ transition is given by
\begin{equation}
P_{e\mu} = \sin^{2}2\theta_{e\mu}\sin^2\left(\frac{\Delta m^2 L}{4 E}\right),
\end{equation}
where $\theta_{e\mu}$ is the effective mixing angle, and $\Delta m^{2}$ is
the effective mass difference, independent of the sterile neutrino
model. In the (3+1) model (with only one sterile neutrino) then
\begin{equation}
\sin^{2}2\theta_{e\mu} \equiv 4|U_{\mu 4}|^2|U_{e4}|^{2}, 
\end{equation}
where $U_{\epsilon n}$ is an element of the PMNS
mixing matrix. The $\nu_\mu$ disappearance probability is 
\begin{equation}
P_{\mu\mu} = 4\left| U_{\mu 4}\right|^2  \left(1-\left| U_{\mu 4}\right|^2 \right)\sin^2\left(\frac{\Delta m^2 L}{4 E}\right).
\end{equation}

\begin{figure}[htbp]
    \includegraphics[width=0.5\columnwidth]{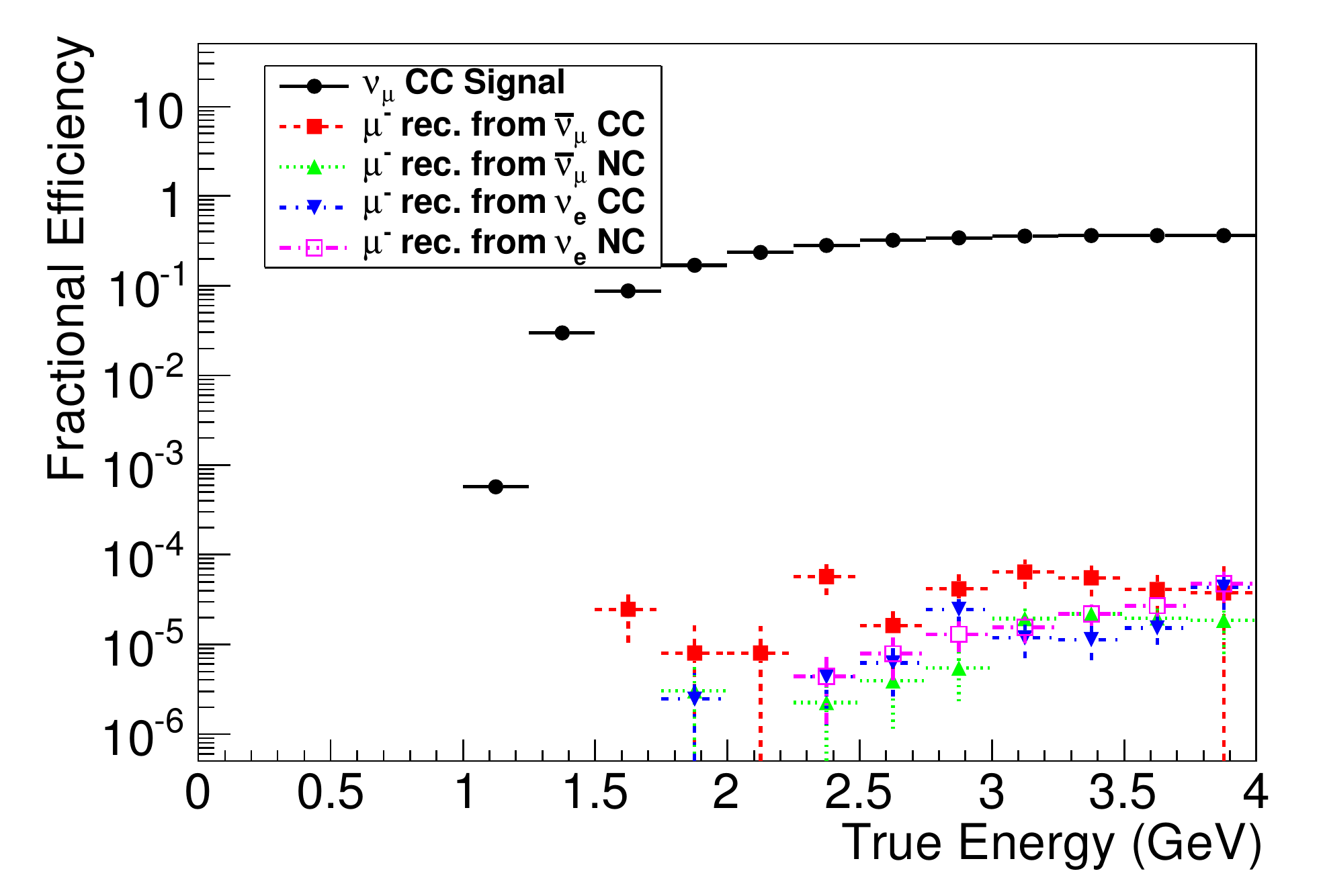}
    \includegraphics[width=0.5\columnwidth]{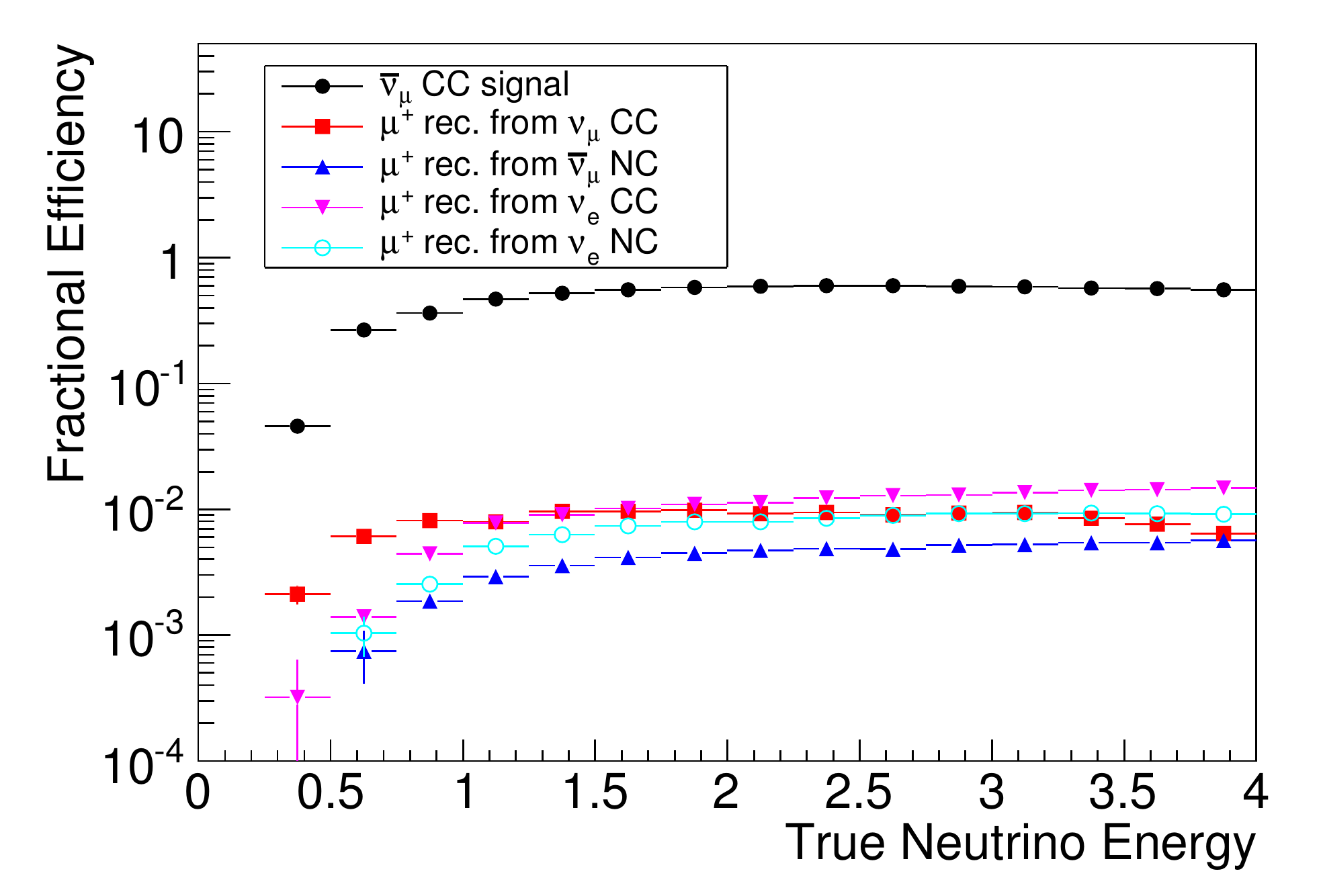}
    \caption{Efficiencies of signals and backgrounds for a $\nu_{\mu}$
    appearance (left) and a $\bar{\nu}_{\mu}$ disappearance (right) neutrino oscillation search at nuSTORM.}
  \label{fig:eff}
\end{figure}

\begin{figure}[htbp]
\includegraphics[width=0.5\columnwidth]{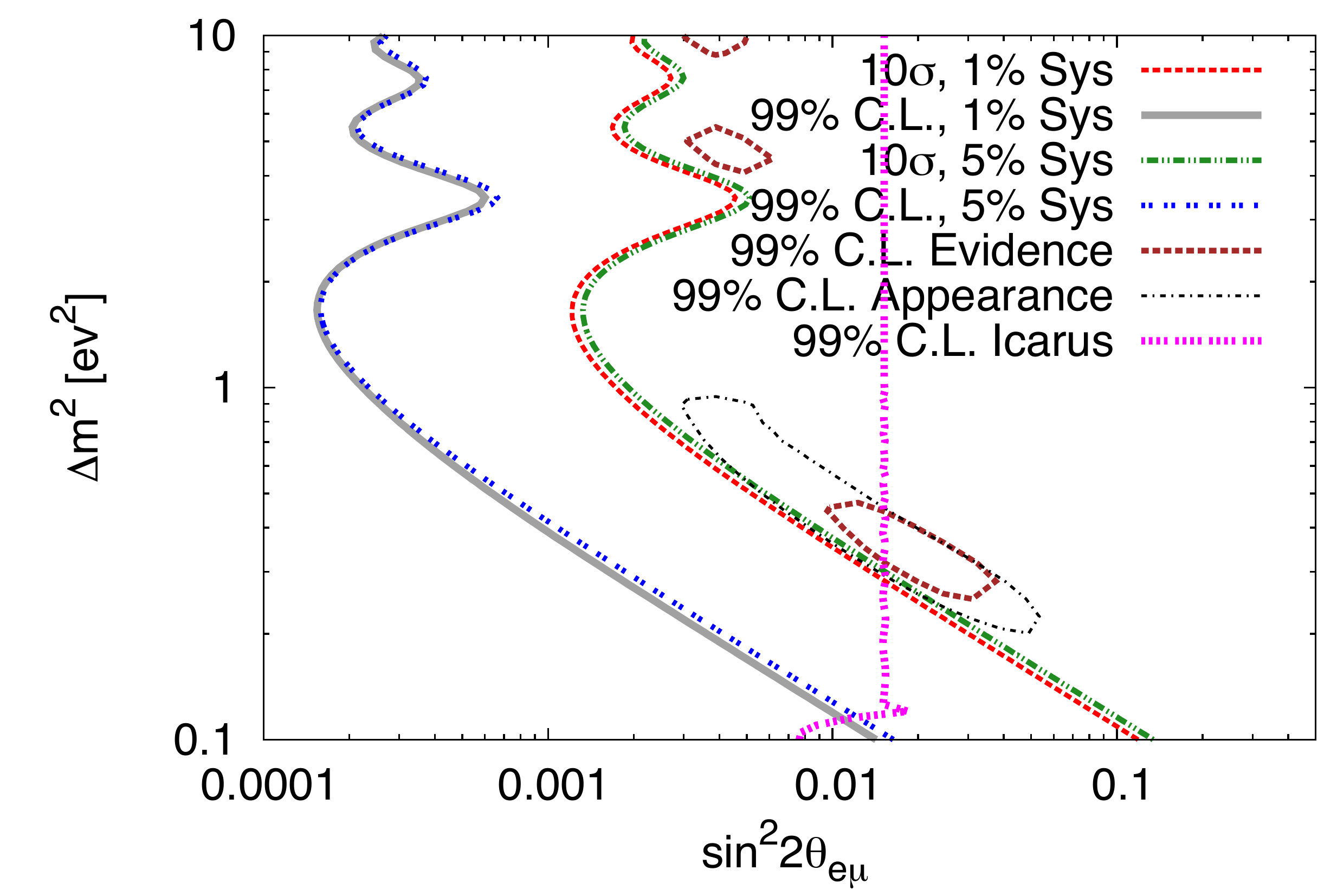}
\includegraphics[width=0.5\columnwidth]{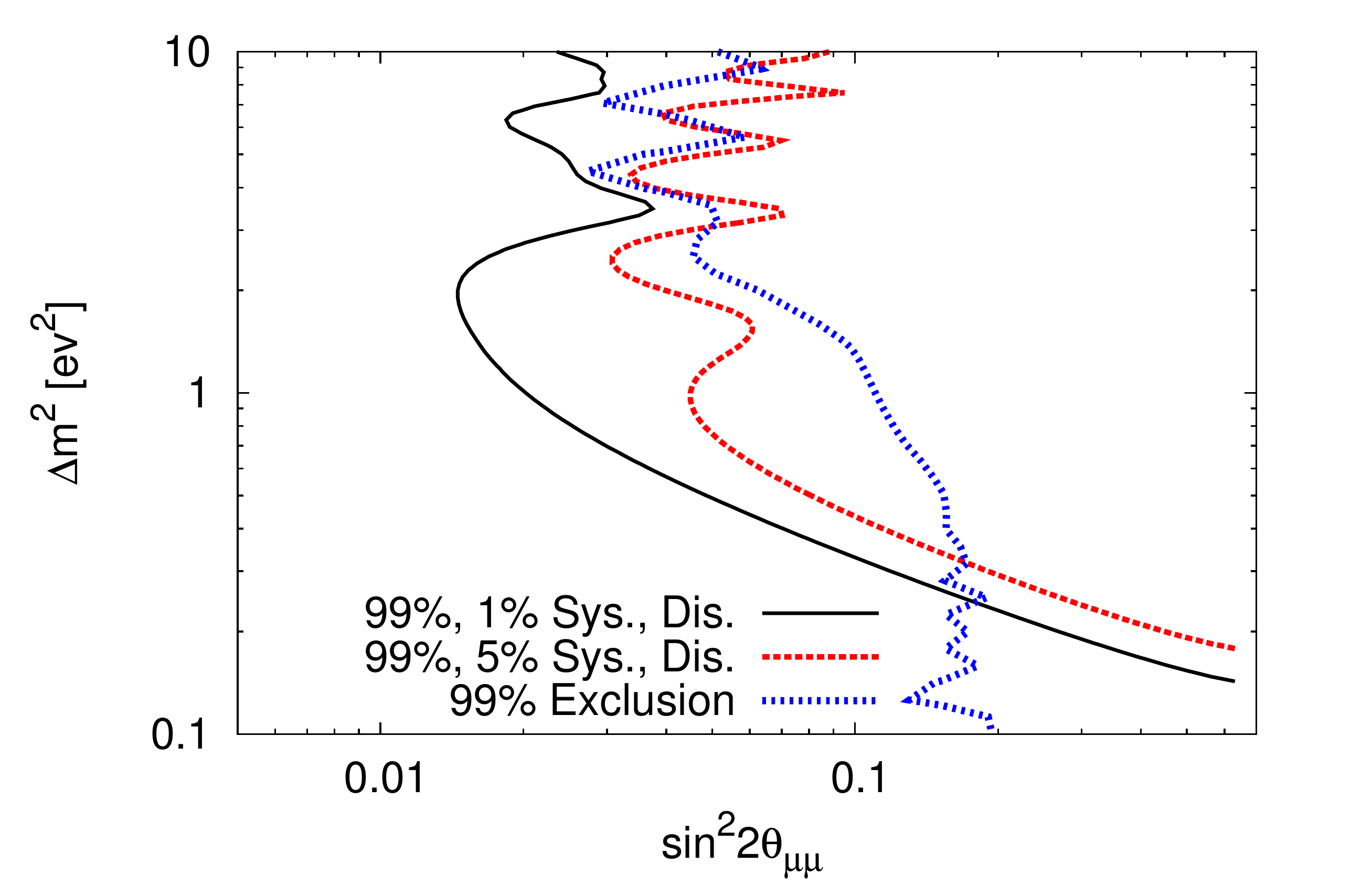}
\caption{Sterile neutrino appearance channel sensitivity (left) at nuSTORM compared to 99\% confidence contours from fits generated by Kopp {\it{ et. al.}} 
\cite{Kopp:2013vaa} and limits set by ICARUS \cite{Antonello:2013gut}. Sensitivity of nuSTORM to $\bar{\nu}_{\mu}$ disappearance (right) oscillations assuming a (3+1) neutrino model, compared to the existing disappearance data \cite{Kopp:2013vaa}.
}
\label{fig:appsig}
\end{figure}

A complete appearance and disappearance analysis was carried out to determine the sensitivity of nuSTORM to search for eV-scale sterile neutrinos \cite{Adey:2014rfv}.
The short-baseline oscillation search was carried out by simulating a near detector at 50 m and a far detector at 2 km, with a $10^{21}$ proton-on-target  exposure. 
The appearance and disappearance analyses were carried out with two optimised multi-variate analyses. The search signal and background efficiencies are shown in Figure \ref{fig:eff} and the sterile neutrino sensitivities also for the appearance and disappearance searches are shown in Figure \ref{fig:appsig}. nuSTORM can either discover or rule out previous sterile neutrino evidence with better than 10$\sigma$ sensitivity.

 \section{Conclusions and Outlook}
The nuSTORM facility  is an entry level Neutrino Factory that can be built now, without need of new technology such as ionisation cooling. It can carry out world-leading
 neutrino scattering physics studies, with the potential to measure all neutrino cross sections with high precision due to the less than 1\% uncertainty on the knowledge of the flux. 
 nuSTORM has the potential to resolve the issue of systematic errors for long-baseline neutrino oscillation experiments searching for CP violation. Hence, the nuSTORM facility would be  the ultimate neutrino scattering physics facility: ``a neutrino light source". 
 
Furthermore, the nuSTORM facility could serve as a test-bed for muon acceleration R\&D, to develop the capabilities of muon storage rings for future particle physics 
projects.  At the end of the first straight, one can place a  3.5~m iron pion absorber. After the absorber, $\sim 10^{10}$ muons per pulse can be observed between 
100--300\,MeV/c. This high intensity muon source can be used to carry out a six-dimensional muon cooling experiment, essential to be able to realise future
Neutrino Factories and Muon Colliders.

The nuSTORM facility could be sited at either Fermilab or in the North Area at CERN. A world-wide neutrino programme in which long-baseline experiments
are carried out at LBNF in the USA and at Hyper-Kamiokande in Japan, could be complemented by the nuSTORM facility, which could be sited either at Fermilab 
or CERN to ensure that maximum impact is delivered from this programme. Therefore, the question should not be whether we can afford nuSTORM, but rather, 
whether we can afford not to have nuSTORM as a facility to perform high precision neutrino studies.

\bigskip

\bibliography{NuPhys_2014_Soler}

\end{document}